\documentclass{emulateapj}   

\usepackage[]{natbib}
\usepackage{epsfig}
\begin{document}

\title{The Biermann Battery in Cosmological MHD Simulations of Population III Star Formation}

\author{Hao Xu\altaffilmark{1,2}, 
Brian W. O'Shea\altaffilmark{2,3}, 
David C. Collins\altaffilmark{1}, 
Michael L. Norman\altaffilmark{1},  
Hui Li\altaffilmark{2}, 
and Shengtai Li\altaffilmark{2}}
\email{haxu@ucsd.edu}
\altaffiltext{1}{Center for Astrophysics and Space Sciences, University of California, 
San Diego, 9500 Gilman Drive, La Jolla, CA 92093}
\altaffiltext{2}{Theoretical division, Los Alamos National Lab, Los Alamos, NM 87545}
\altaffiltext{3}{Department of Physics and Astronomy and Lyman Briggs College, Michigan 
State University, East Lansing, MI 48824}

\begin{abstract}
We report the results of the first self-consistent three-dimensional adaptive mesh refinement 
magnetohydrodynamical simulations of Population III star formation including the 
Biermann Battery effect.  We find that the Population III stars formed including this 
effect are both qualitatively and quantitatively similar to those from hydrodynamics-only 
(non-MHD) cosmological simulations.  We observe peak magnetic fields of 
$\simeq 10^{-9}$~G in the center of our star-forming halo at $z\simeq 17.55$.  
The magnetic fields created by the 
Biermann Battery effect are predominantly formed early in the evolution of the primordial
halo at low density and large spatial scales, and then grow through compression
 and by shear flows.  The fields seen in this calculation are
never large enough to be dynamically important (with $\beta \geq 10^{15}$ at all times),
and should be considered 
the minimum possible fields in existence during Population III star formation, and may 
be seed fields for the stellar dynamo or the magnetorotational instability at higher 
densities and smaller spatial scales.
\end{abstract}


\maketitle


\section{Introduction}
The nature of the first generation of stars, as well as their influence on later
structure formation, is a fundamental problem in cosmology.  A
great deal of theoretical progress has been made (see recent reviews by 
\citet{Bromm04},~\citet{2005SSRv..117..445G}, and~\citet{Ciardi05}.  In the past decade, 
cosmological hydrodynamic simulations
of Population III star formation have achieved great success, and significantly
different numerical methods have produced results that agree quite well
\citep{Abel02, 2002ApJ...564...23B,Yoshida03, O'Shea07}.  These calculations have given
 a reasonably clear
picture of the formation process of Population III stars, and have provided some constraints on
 many of their important properties.

These calculations, while useful, largely ignore an important issue: the relevance of magnetic
fields in Population III star formation.  Magnetic fields are widely observed in our galaxy, 
in other galaxies, and in galaxy clusters, and the origin of these fields is one of
the most fundamental and challenging problems in astrophysics \citep{Carilli02, Widrow02}.
One possibility is that magnetic fields are created and amplified in the first generation
of stars and are spread throughout the IGM when these stars explode, providing seed
fields for later generations of stars and for further amplification by dynamo effects.  If some 
seed magnetic fields exist before Population III stars form, they may help to remove 
angular momentum from the star-forming clouds, significantly changing the ultimate
mass range of these stars \citep{Pudritz89, Davies00}.

Several groups have examined the importance of magnetic fields on the evolution of 
Population III protostellar disks using analytic or semi-analytic models.  These
include 
\citet{silk06}, \citet{tan04a} and \citet{tan04b}, who model (among other aspects of
primordial star formation and evolution) dynamos in primordial accretion disks.  Other
authors, including \citet{flower03} and \citet{maki07}, use one-zone calculations
to examine the collapse of the 
primordial star-forming cloud and the assumptions of flux-freezing.  While useful,
these models do not self-consistently include the effects of both magnetic fields and
cosmological structure formation.

In this Letter, we report the results of the first
magnetohydrodynamic simulations of Population III
star formation including the Biermann Battery effect \citep{Biermann50} within the
context of cosmological structure formation.  We describe
our numerical methods and the simulation used in Section~\ref{sec:methods}, present
our key results in Section~\ref{sec:results}, and discuss and summarize our
results in Section~\ref{sec:discussion}.

\section{Methodology}
\label{sec:methods}

The simulation discussed in this Letter was performed using 
the adaptive mesh N-body plus hydrodynamics code Enzo \citep{oshea04,norman07}, which
has recently been extended to including the equations of ideal magnetohydrodynamics.
The MHD solver is a high-order Godunov-type finite-volume numerical solver 
\citep{Li03, Li05, Li08}. This solver was recently successfully
 used to study magnetic jets  \citep{Li06,Nakamura06,Nakamura07}. To maintain the
divergence-free magnetic field condition, we used a constrained transport (CT) scheme  
\citep{Balsara99} as well as a modified divergence-free reconstruction scheme 
original proposed by \citet{Balsara01} that includes second-order-accurate divergence-free 
restriction and prolongation for magnetic fields in the adaptive mesh hierarchy \citep{Collins08}. 
These equations are solved in cosmological calculations in the comoving frame, as described 
in \citet{2008ApJ...681L..61X}.
 To track the pressure more accurately in hypersonic regions, we have implemented the 
modified entropy equation given in \citet{Ryu93} and the internal energy equation 
given in \citet{Bryan95} in our code.

To study the effect of the Biermann battery during the formation of Population III stars, 
the battery term is added to the induction equation~\citep{Kulsrud05}:

\begin{eqnarray}
\nonumber
\frac{\partial \bf{B}}{\partial t} & = & \nabla \times (\bf{v} \times \bf{B}) 
+ \frac{ c \nabla p_e \times \nabla n_e}{n_e^2 e}  \\
\nonumber
                                   & = & \nabla \times (\bf{v} \times \bf{B}) 
+ \frac{c m}{e} \frac{1}{1+\chi} \frac{\nabla p \times \nabla \rho}{\rho^2}  {\mathrm,}
\end{eqnarray}

where $c$ is speed of light, $p_e$ is the electron pressure, $n_e$ is the electron 
number density, $e$ is the electron charge, $m$ is the average mass per particle and 
$\chi$ is the ionization fraction. This term is non-zero in regions where curved shocks
are formed. The second equation assumes that $m$ and $\chi$ are unchanged in space, 
which is a reasonable approximation locally. As pointed out by \citet{Kulsrud97}, 
the ionization fraction enters into the induction equation through $1+\chi$, so even a 
very small ionization fraction is enough to generate magnetic fields. In our simulation, 
we took $m=1.2$ $m_H$, where $m_H$ is the mass of a 
neutral hydrogen atom.   Note that the electron fraction is $\ll 1$ at all times in this
star formation scenario, so we set $\chi = 0$ rather than calculating the electron fraction in
every cell.  Given the approximation above, this introduces negligible error.
The battery term is added to the simulation 
through the EMF term in the constrained transport algorithm, ensuring that no divergence of magnetic 
fields is introduced to the system. 

Since the electron fraction 
declines rapidly at the halo core, ambipolar diffusion may be important in star formation 
simulation \citep{Abel02}. To address this issue, we compared the ambipolar diffusion timescale 
with the dynamic timescale in our simulation, as was done by \citep{O'Shea07}. The ratio is 
always larger than 10 over the density range of interest, suggesting that ambipolar diffusion 
does not play a significant role up to the densities that we study.

Our simulation setup is similar to that described in \citet{O'Shea07}.  We use a N-body
plus hydrodynamics 
simulation with a comoving box size of 0.3 h$^{-1}$ Mpc, initialized at $z=99$ with an 
Eisenstein \& Hu power spectrum~\citep{eishu99}, and with cosmological 
parameters $\Omega_b = 0.04$, $\Omega_m = 0.3$, $\Omega_\Lambda = 0.7$, $h=0.7$, 
$\sigma_8 = 0.9$, and $n_s = 1.0$.  The simulation was first run with a $128^3$ root
grid and three levels of AMR to $z=15$, where the most massive halo in the volume
was found.  The simulation was then re-centered on this halo, using a $128^3$ root
grid and three levels of static nested grids, giving a dark matter and baryon resolution
of $1.81$~h$^{-1}$~M$_\odot$ and $0.28$~h$^{-1}$~M$_\odot$, respectively, and an
initial comoving spatial resolution of 293~h$^{-1}$~pc.  The simulation was
started at $z=99$, initialized with zero magnetic fields throughout,
 and allowed to run with a maximum of 22 levels of adaptive mesh 
refinement until the collapse of the most massive halo, at $z \simeq 17.55$, using
the nine-species molecular chemistry of \citet{abel97} and \citet{anninos97}, but
modified for high densities as described in \citet{Abel02}. Further details
of the simulation setup and physics can be found in \citet{O'Shea07}.

\section{Results }
\label{sec:results}

\begin{figure}
\includegraphics[width=0.4\textwidth]{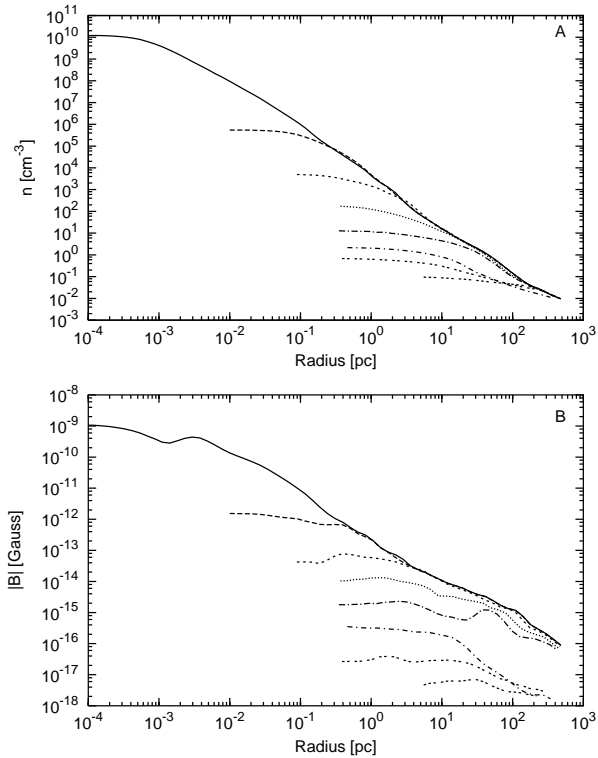}
\caption{Evolution of spherically-averaged, mass-weighted radial profiles 
of baryon number density (top) and magnetic field strength (bottom) of 
the Population III star-forming halo.  Lines correspond to
(from bottom to top in each panel) $z = 40, 30, 25, 20, 19, 18, 17.61, 17.55$.
} 
\label{fig:fig1}
\end{figure}

\begin{figure}
\includegraphics[width=0.4\textwidth]{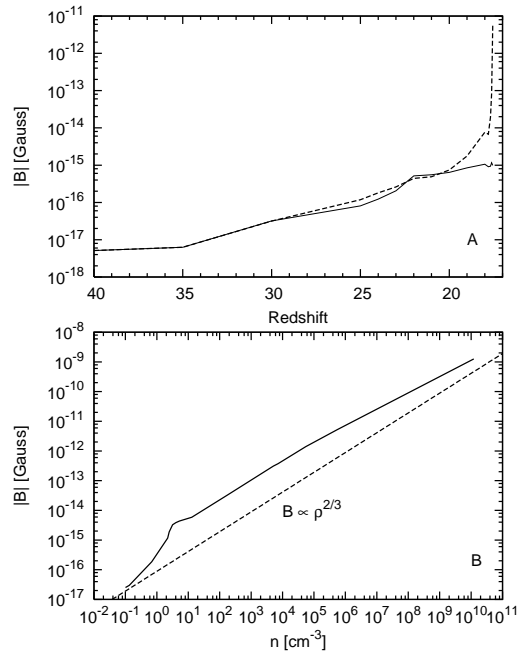}
\caption{Top panel:  mass-weighted (dashed line) and volume-weighted (solid line)
mean magnetic field strength inside the virial radius of the collapsing
halo as a function of redshift.  Bottom panel: mass-weighted mean magnetic field
strength as a function of baryon number density inside the core of the primordial 
star forming halo at the final simulation output 
(solid line; M$_{enc} \simeq 100$~M$_\odot$). The
dashed line shows $|B| \propto \rho^{2/3}$ (as expected by flux freezing 
and spherical collapse).
}
\label{fig:fig2}
\end{figure}

Over the course of the simulation, magnetic fields are produced at
various physical scales by the Biermann Battery process.  Figure \ref{fig:fig1} shows the
evolution of spherically-averaged radial profiles of the baryon number density and magnetic 
field strength of the primordial star-forming halo at several times during
the halo evolution.  Since the magnetic fields are very small 
($\beta \equiv P_{thermal} / P_{magnetic}$ is greater than $10^{15}$ at all times), 
the gas collapse at the halo core is almost the same as the hydrodynamic
($B = 0$) case described by \citet{O'Shea07}. The profiles start at $z=40$, 
where the magnetic fields generated by the Biermann Battery are of order $10^{-18}\ G$, 
which is consistent with theoretical predictions \citep{Pudritz89, Davies00}. 
In plasmas with a large beta ($\gg 1$) and large mass-to-flux ratio, magnetic
fields follow the gas 
falling into the core passively, so after being generated at low densities, the
 magnetic fields are primarily amplified by being carried along with collapsing
baryons.  This is shown more clearly in Figure~\ref{fig:fig2}, and will be
discussed in more detail later.
At low density (up to n$_b \simeq 2$~cm$^{-3}$, corresponding
to the evolution of the gas up to $z \sim 20$), the Biermann Battery is effective at
creating magnetic fields.  The central magnetic field scales with density 
as $|B| \propto \rho$, suggesting
that the magnetic field is enhanced by both the Biermann Battery effect
and by gas collapse.  At higher densities (after $z \simeq 20$), the magnetic field
at the center of the halo is amplified as $|B| \propto \rho^{2/3}$, as expected in
spherical collapse.  At later times,
when the gas in the center of the halo collapses, the magnetic field in the center of
the halo rises to $\sim 10^{-9}$~G as the central number density grows to
n$_b \simeq 10^{10}$~cm$^{-3}$.

It is useful to examine the history of magnetic field amplification inside the 
 halo in more detail.  This is shown in  
Figure~\ref{fig:fig2}, which shows the mass- and volume-weighted magnetic
fields inside the halo virial radius as a function of redshift (top panel), and also
shows the mass-weighted mean magnetic field strength in the inner
$100$~M$_\odot$ of gas as a function of the mass-weighted baryon
number density of gas in the same Lagrangian region (bottom panel).
From these two plots, the two stages of magnetic field amplification
can be observed.  From $z=40$ to $z=20$, the magnetic fields slowly
increase to $10^{-15}$~G.  At this stage,
the mass-weighted and volume-weighted fields are quite similar, because
the magnetic fields are created and amplified by the Biermann Battery
at large spatial scales, and the fields are distributed uniformly by 
mergers and shear flows (see \citet{Roettiger99}).  During
the second stage of evolution, after $z=20$, the volume-weighted and 
mass-weighted fields evolve in very different ways, with the mass-weighted
magnetic field growing rapidly along with the increasing baryon density in the
halo cores.

The bottom panel of Figure~\ref{fig:fig2} shows that at
late times (high densities), the magnetic
field in the core of the halo increases as $|B| \propto \rho^{2/3}$, or
solely from field amplification due to magnetic fields frozen into
spherically-contracting gas.  The collapse timescale at the center of the halo is proportional
to the cooling time, which is quite short at high densities, making
any significant magnetic field creation by the Biermann Battery mechanism
difficult in the halo center.
The baryon density in the halo core grows about 9 orders of magnitude
(from 1 to $10^{10}$~cm$^{-3}$ between $z=20$ and $z=17.55$, when the 
simulation is terminated), and the magnetic fields in the core increase by 
about 5 orders of magnitude (from $10^{-15}$ to $10^{-9}$ G).  Though the 
core region contains less than one percent of the halo gas, the mass-weighted 
magnetic field in the halo increases by approximately four orders of magnitude,
from $\sim 10^{-15}$ to $10^{-11}\ G$, between z=20 and z=17.55.                  

\begin{figure}
\includegraphics[width=0.4\textwidth]{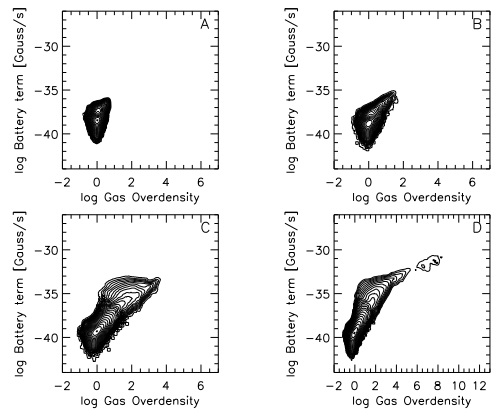}
\caption{Mass-weighted two-dimensional distributions of the instantaneous rate
of magnetic field generation via the Biermann battery 
vs. baryon overdensity, at $z=40$ (top left), $z=30$ (top right), 
$z=20$ (bottom left) and $z=17.55$ (bottom right).  Each panel includes
all gas within the virial radius at a given redshift. Contour lines are
logarithmic, and show relative values between the highest and lowest 
non-zero values in each panel.}
\label{fig:magrho}
\end{figure}

To show where magnetic field generation via the Biermann Battery takes place,
and to provide a more quantitative understanding of its effect in different
stages of halo evolution, we show in Figure~\ref{fig:magrho}
the two-dimensional distribution of the instantaneous magnetic field 
generation rate via the Biermann Battery versus baryon overdensity 
at four  simulation outputs from $z=40$ to $z=17.55$.  
At early times, very
little gas is at high overdensities, and the magnetic field generation
rate is below $10^{-35}$~G/s. 
The total time taken during the simulation, from $z=99$ to
$z=17.55$, is approximately $6.2 \times 10^{15}$ seconds.  If the
magnetic fields are only created by the Biermann Battery without
any amplification from shear flows or collapse, the magnetic fields
would be at most $10^{-19}$~G at $z \sim 18$.  By $z=40$ 
($\sim 1.5 \times 10^{15}$ seconds, or $5 \times 10^7$ years, 
after the beginning of the calculation),
fields reach a strength of $10^{-18}$ G, suggesting that
some amplification is taking place through gas collapse.
While the rate of magnetic field generation via the Biermann Battery can
actually be much higher at high overdensity (as seen in the bottom right
panel, at $z=17.55$), the rate never exceeds $10^{-30}$~G/s 
($\sim 3 \times 10^{-23}$~G/yr). Given that
the collapse of the core, from a peak baryon density
of n$_b \sim 1$~cm$^{-3}$ to $10^{10}$~cm$^{-3}$, takes approximately $10^{14}$
seconds ($\simeq 3 \times 10^6$ years), magnetic fields larger 
than $10^{-16}$~G cannot be made with the Biermann
Battery during this time, strongly suggesting that the battery effect 
makes little contribution to magnetic field amplification during the final
collapse.

\section{Discussion and Conclusions}
\label{sec:discussion}

In this Letter, we have shown results from the first cosmological
magnetohydrodynamical simulation of
Population III star formation including the Biermann Battery effect.  
This effect is one of the most robust methods of generating magnetic
fields in the Universe \citep{Biermann50}, and thus provides useful constraints
on the minimum magnetic field expected in situations where Population III star formation
will take place. 

We find from our simulation that small magnetic fields are primarily generated
via the Biermann Battery at relatively low overdensities, and are then amplified 
to values of nearly $10^{-9}$~G at the center of the cosmological halo 
 via gravitational collapse.  While significant, this magnetic field
is still quite small -- the plasma $\beta$ is never smaller than $10^{15}$ at any point 
during the simulation.  This suggests strongly that the magnetic fields
do not play a significant dynamical role up to n$_B \sim 10^{10}$~cm$^{-3}$, when
the simulation is terminated.  As a result, the evolution of the primordial gas
within the cosmological halo is very close to the hydrodynamic results obtained
by \citet{O'Shea07}. At later times, however, the magnetic fields may be amplified
to dynamically relevant values via a process such as the magnetorotational
instability or a dynamo
 \citep[][]{silk06,tan04a,tan04b}.

The simulation of further evolution of our evolving protostellar core is
currently beyond the capability of our numerical tools.  However, there are
multiple avenues that we can follow to more thoroughly explore the 
relevance of magnetic fields in Population III star formation.  We
can begin our calculation with seed fields having strengths based on limits
from observation and theory~\citep{Widrow02}.  We can examine the
effects of magnetic fields at higher densities by implementing more
 chemical and cooling processes \citep[as discussed by ][]{turk08}.
Both of these projects are underway, and we will report on the 
results in an upcoming paper.

To summarize, we have performed the first calculations that incorporate
the Biermann Battery in cosmological magnetohydrodynamic simulations of
Population III star formation.  Our key results are as follows:

1.  From an initial state with no magnetic fields, a combination of
the Biermann Battery and compressional amplification can result in fields 
with strengths of $|B| \simeq 10^{-9}$~G at n$_B \simeq 10^{10}$~cm$^{-3}$
at the center of a cosmological halo where a Population III star will form.

2.  The Biermann Battery creates fields predominantly at low density
(n$_B \leq 10$~cm$^{-3}$) and large spatial scales in Population III
star-forming halos.

3.  The magnetic fields created by the Biermann Battery are dynamically
unimportant at all densities below n$_B \simeq 10^{10}$~cm$^{-3}$ --
$\beta \equiv P_{th} / P_{B} \geq 10^{15}$ at all times during the 
evolution of the halo.


\acknowledgments{
We would like to thank Tom Abel, Greg Bryan, Mordecai-Mark Mac Low,
Matthew Turk, and Dan Whalen
for useful conversations.
HX and BWO are supported by IGPP at Los Alamos National Laboratory,
and carried out this work under the auspices of the
National Nuclear Security Administration of the
U.S. Department of Energy at Los Alamos National
Laboratory under Contract No. DE-AC52-06NA25396.  
HX and MLN acknowledge partial support from NSF grant AST-0708960.
The simulation described in this paper was performed at 
the San Diego Supercomputing Center with computing time provided by 
NRAC allocation MCA98N020.
}

\bibliographystyle{apj}
\bibliography{apj-jour,mybib}  

\end{document}